\documentclass[reprint,amsmath,amssymb,aip,jcp]{revtex4-1} 	

\usepackage{graphicx,color}	
\usepackage[dvipsnames]{xcolor}						
\usepackage{amssymb}
\usepackage{natbib}
\usepackage{array}
\usepackage{subfigure}

\newcommand{\ft}[1]{{\color{black} #1}}

\begin{document}


\title{Structural Covariance in the Hard Sphere Fluid}

\author{Benjamin M.G.D. Carter}
\email{benjamin.carter@bristol.ac.uk}
\affiliation{HH Wills Physics Laboratory, Tyndall Avenue, Bristol, BS8 1TL, UK.}
\author{Francesco Turci}
\affiliation{HH Wills Physics Laboratory, Tyndall Avenue, Bristol, BS8 1TL, UK.}
\author{Pierre Ronceray}
\affiliation{Princeton Center for Theoretical Science, Princeton University, Princeton, NJ 08544, USA}
\author{C. Patrick Royall} 
\affiliation{HH Wills Physics Laboratory, Tyndall Avenue, Bristol, BS8 1TL, UK.}
\affiliation{School of Chemistry, University of Bristol, Cantock Close, Bristol, BS8 1TS, UK.}
\affiliation{Centre for Nanoscience and Quantum Information, Tyndall Avenue, Bristol, BS8 1FD, UK.}

\begin{abstract}
We study the joint variability of structural information in a hard sphere fluid biased to avoid crystallisation and form fivefold symmetric geometric motifs. We show that the structural covariance matrix approach, originally proposed for on-lattice liquids \ft{[Ronceray and Harrowell, JCP 2016]}, can be meaningfully employed to understand structural relationships between different motifs and can \textit{predict}, within the linear-response regime, structural changes related to motifs distinct from that used to bias the system.
\end{abstract}

\maketitle

\section{Introduction}
Short-range local order is a distinctive feature of the liquid state. In models of simple liquids such as the Lennard-Jones liquid or the hard sphere fluid, local structure has been studied via the measurement of pair correlation functions (which define a characteristic correlation length) \cite{hansen} or with higher order correlations, such as rings of particles and recurrent geometric motifs, since the early times of the theory of liquids, with the pioneering work of Bernal \cite{bernal1959,bernal1960} and Finney \cite{finney1970, finney1970mc} in ``ball-bearing'' models.

Since then, more sophisticated probing techniques have been developed to characterise the local structure of disordered systems: projection of the nearest neighbours onto spherical harmonics \cite{Steinhardt1981,lechner2009}; the statistics of Voronoi polyhedra and their facets \cite{coslovich2007}; the analysis of common neighbours \cite{honeycutt1987}; the match of local motifs with minimum energy clusters \cite{malins2013tcc}; persistence homology of rings of particles \cite{hiraoka2016} are just a few examples.

The idea underpinning these analyses is that the knowledge of the degree of local order may shed light on interesting dynamical and thermodynamical properties of disordered systems in general and of liquids in particular. These include possible signatures of precursors to crystallisation in metastable liquids \cite{schilling2010, russo2012} as well as the eventual coupling between structural and dynamical heterogeneities in supercooled liquids and glasses (for a review on structure in dynamically arrested systems see \cite{royall2015physrep}).

A major issue in this approach is the fact that different diagnostic and analysis tools of local structural properties may lead to different conclusions on the role of local structure in liquids. For example, the role of crystalline and icosahedral order in supercooled liquids has been extensively debated \cite{finney1970,medvedev1987,anikeenko2007,coslovich2007, leocmach2012,royall2015} and the metrics used to determine each of those orders play a role in the interpretation of the results. Understanding how different types of local structural motifs correlate would permit us to systematically compare different metrics, and thus open the way towards a unified quantitative framework for local liquid order.

Here we consider the problem of the classification of local order in a canonical liquid from a simple statistical point of view. Recent work on a toy model of a lattice liquid with a purely structural energy landscape, the Favoured Local Structures model\cite{ronceray_favoured_2015}, has demonstrated the importance of correlations between different structural geometric motifs present in the liquid. Indeed, the statistics of high-temperature structural fluctuations provides key information on the liquid entropy \cite{ronceray_geometry_2012}, while their correlations provides  a quantitative metrics for the stability/instability of the liquid towards crystal formation, being good predictors of crystallisation times and surface tensions \cite{ronceray_liquid_2016,ronceray2017}. Inspired by these results on a highly idealised system, we study here the structural statistics in the hard sphere fluid at high packing fraction, a much more realistic liquid model. Following closely the approach proposed by Ronceray and Harrowell \cite{ronceray_liquid_2016}, we measure \textit{structural covariances} and show how they encode, at the same time, geometric information on the classification itself and physical information on the propensity of the system to form crystalline or fivefold symmetric structures. 

The article is organised as follows: in Section \ref{sec:hardsph} we introduce the studied model and the structural classification of reference; in Section \ref{sec:structcov} we discuss the structural covariance formalism and its main results in the case of hard spheres; in Section \ref{sec:linear} demonstrate that the covariance framework allows us to predict quantitatively the parameter dependence of the liquid structure; and in  Section \ref{sec:conclusions} we summarise our findings and propose further directions of research.

\section{Hard spheres with structural bias}
\label{sec:hardsph}
 In the hard-sphere liquid, fivefold symmetry plays an important role, frustrating the formation of crystalline order \cite{charbonneau2012,royall2014angell,taffs2016}. The degree of fivefold frustration is often quantified in terms of the number of fivefold symmetric structures, identified through the pentagonal bipyramid, a geometrical arrangement which is formed by a bonded spindle pair of particles sharing exactly five neighbours. 

In order to study the local structure of the system, including fivefold symmetry, we employ the Topological Cluster Classification (TCC) \cite{malins2013tcc}. This algorithm has been successfully used in the past to study the structure of simple liquids \cite{malins2013isomorph}, gels \cite{royall2015prl,razali2017, griffiths2017}, glasses \cite{malins2013fara, malins2013jcp} and athermal packings \cite{liu2015adhesive}. It identifies a total of 33 structures based on minimum energy clusters of elementary pair potentials, such as the Lennard-Jones and Morse liquids. Its labelling of different structures is inherited from the labelling of minimum energy clusters of simple liquids (with Lennard-Jones, Morse or Dzugutov interactions) in the Cambridge Cluster Database \cite{cambridgeDatabase}. Labels are typically formed by a number and a letter: the former refers to the number of particles in the motif, the latter indicates the nature of the potential the motif is a minimum of (letters from A to F correspond to the Morse potential with increasing range, Z stands for the Dzugutov potential \cite{dzugutov1991}, K and W stand for particular forms of the Lennard-Jones potential and X for a BCC crystalline arrangement)\cite{doye1995}. 

\begin{figure*}[t]
\begin{center}
\includegraphics[scale=1]{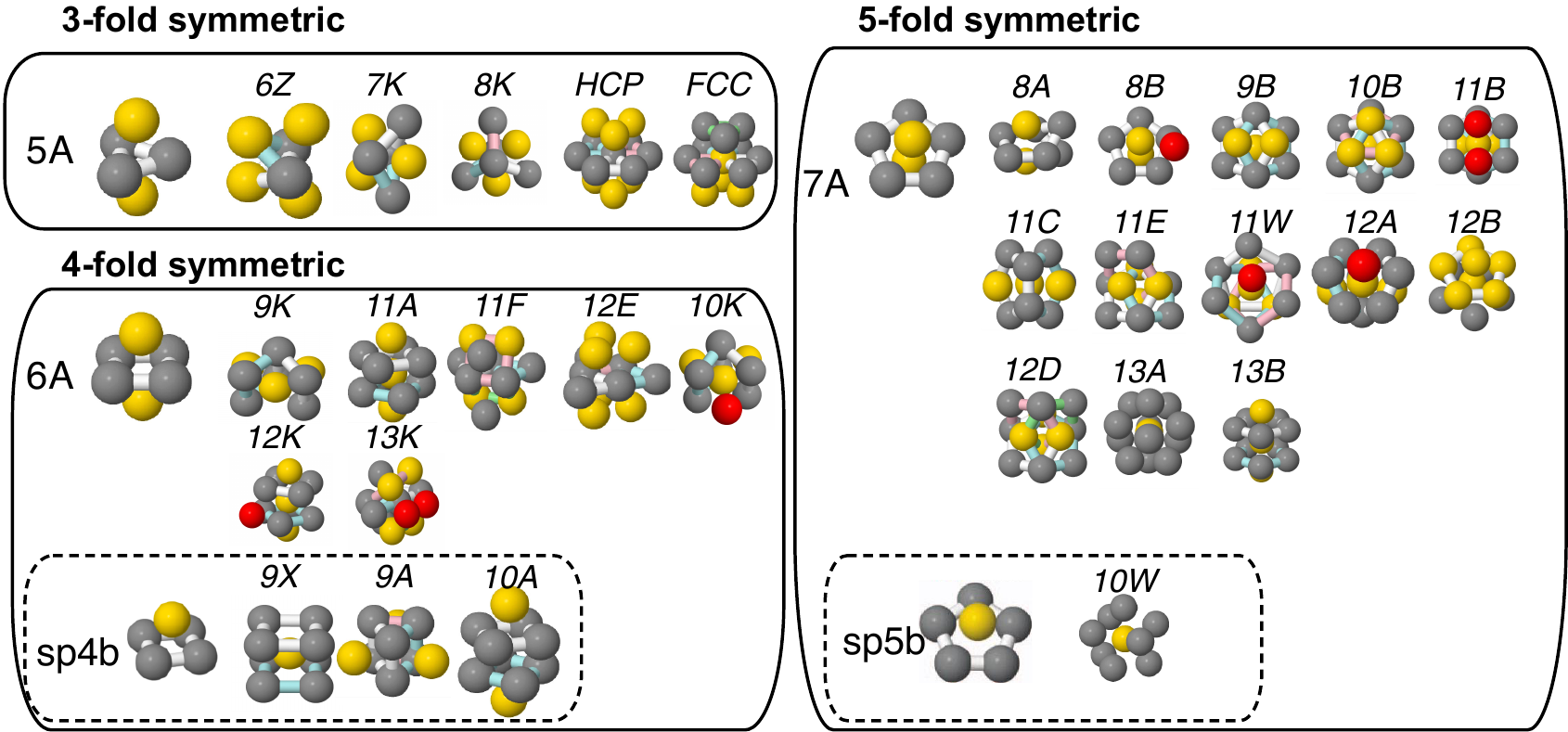}
\end{center}
\caption{\label{FigFamilies}\ft{Topological cluster classification. (Left) Structural motifs related to threefold symmetric ($5A$) or four-fold symmetric ($6A$ and sp4b) local order. The sp4b unit is smaller than the octahedron and  three small motifs are derived from it; its subgroup is highlighted by a dashed line. Rings are represented by colored sticks connecting grey particles, spindle particles are in yellow and additional particles are in red, as in Reference \cite{malins2013tcc}. (Right) Structural motifs related to pentagonal ($7A$ or sp5b) local order.  The sp5b unit is smaller than the pentagonal bi-pyramid and only one motif is derived from it, its subgroup highlighted by a dashed line. Notice the presence of multiple interlaced pentagonal rings in the larger structures such as 10B or 11E. }}
\end{figure*}

In Fig.~\ref{FigFamilies} we illustrate the relationship between the several structures defined in the TCC. We differentiate the several families of structures present in the classification: three-fold (tetrahedral), four-fold and five-fold symmetric structures of different numbers of particles are defined. In particular, the pentagonal bipyramid is termed ``7A''. Hence, we define the total number of pentagonal bipyramids as $N_{7A}$. In this classification, small structures can be part of larger structures. Such a multiple counting contributes to the total number $N_i$ of structures of a given type $i$. In contrast with previous studies \cite{malins2013tcc}, $N_{7A}$ does not correspond to the number of particles detected in pentagonal bipyramids, but to the actual number of (possibly overlapping) bipyramids detected in the liquid, and similarly for all other structures $i$. The relation between the number of bipyramids and the probability to find a particle in a bipyramid is nontrivial, since particles can be part of several overlapping bipyramids.

\begin{figure}[t]
\begin{center}
\includegraphics[scale=1]{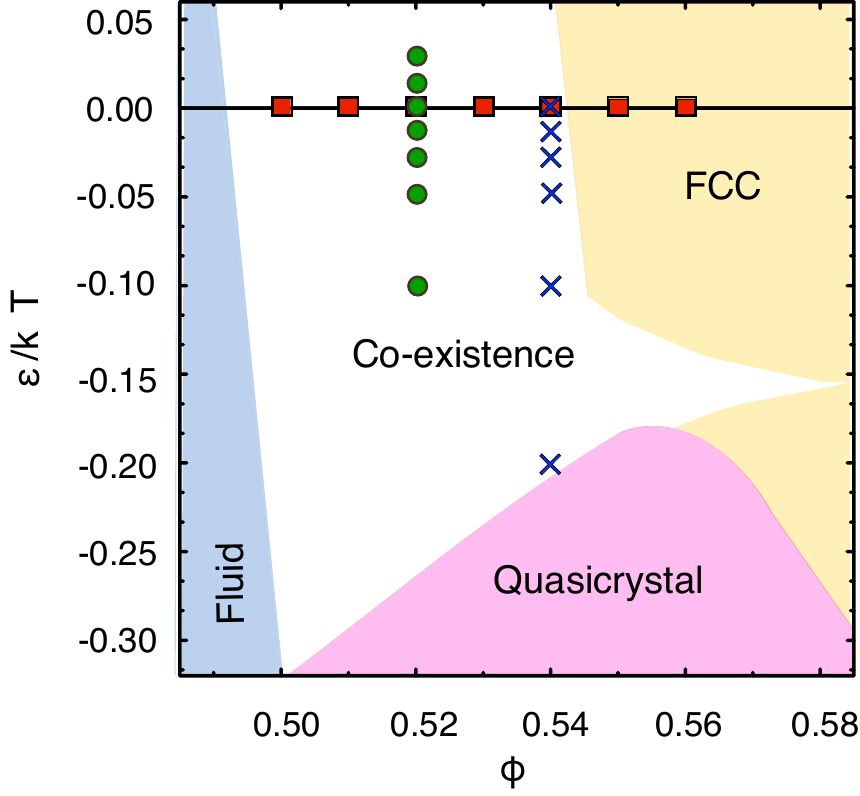}
\end{center}
\caption{\label{FigPhaseDiagramSchematic} Phase diagram of biased hard spheres at high packing fraction. We explore several state points: at zero bias (red squares) with packing fractions $\phi\in [0.52, 0.56]$ and t fixed packing fraction and variable bias, $\phi=0.52$, $\varepsilon\in[-0.10,0.03]$ (green circles) and $\phi=0.54$, $\varepsilon\in[-0.20,0]$ (blue crosses). Phase boundaries are reproduced from \cite{taffs2016}.}
\end{figure}

 In Reference \cite{taffs2016}, fivefold symmetry in hard spheres has been studied through the addition of a many-body energy term $H_{\rm fivefold} =\varepsilon N_{7A}$ to the Hamiltonian of the system, sampling via Monte-Carlo an extended two-dimensional phase diagram in the packing fraction $\phi$ and bias energy $\varepsilon$ (with unit temperature), see Fig.~\ref{FigPhaseDiagramSchematic}. This model exhibits a rich phase behaviour: biasing the system to more negative/positive values of $\varepsilon$  pushes the fluid-solid phase transition to higher/lower packing fraction; at strong enough biases, the system spontaneously nucleates a quasi-crystalline phase rich in five-fold symmetric icosahedra. We refer the reader to Reference\cite{taffs2016} for a more complete discussion of the phase behaviour of the 7A-biased hard-sphere fluid.

In the present article, we extend this work and re-examine runs of $N=2048$ hard spheres in the isothermal-isochoric ensemble for different values of $\phi$, with a specific interest in the influence of the fivefold bias $\varepsilon$ on the structure of the liquid. This parameter fully determines the Hamiltonian of the system, as the hard-sphere interaction has no other contribution than forbidding configurations with overlaps. This model is thus entirely specified by a simple local energy landscape, making it ideally suited for a first study of \emph{structural covariance} in off-lattice systems.

\section{Structural covariance formalism}
\label{sec:structcov}

At any given time, the number $N_i(t)$ of structures of type $i$ in
the system will exhibit some deviation to its mean, reflecting the
randomness of the configurations.  The keystone of our statistical
analysis of liquid structure is the covariance matrix $C_{i,j}$
between these numbers $N_i$ and $N_j$ of structures of types $i$ and
$j$, which reflects the correlations between these random
variables. We now explain how we compute this matrix, before
discussing its structure.

\begin{figure}[t]
  \includegraphics[scale=1]{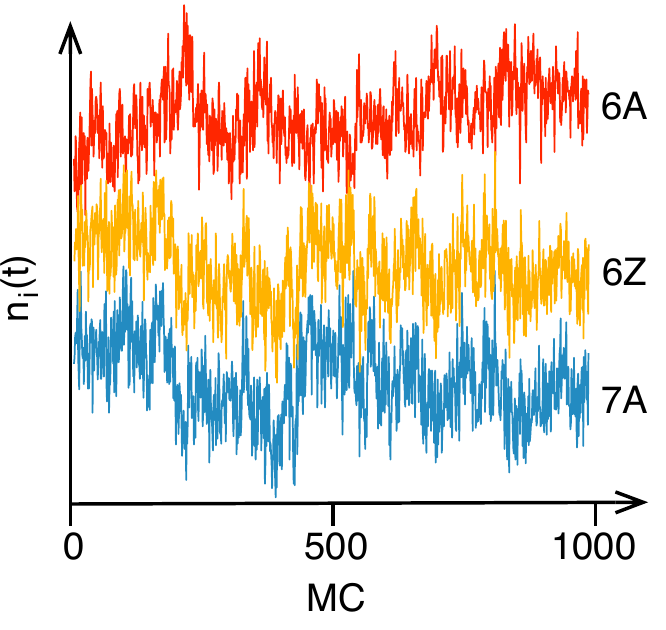}
  \caption{Time evolution in Monte-Carlo sweeps (MCs) of the concentration $n_i$ for the four-fold symmetric $6A$, three-fold symmetric $7A$ and the five-fold symmetric $7A$. Concentration $n_i$ are rescaled and shifted to more visually highlight time correlations (and anticorrelations) between the different time signals.}
  \label{FigTimeSeries}
\end{figure}

We consider Monte-Carlo simulations of biased and unbiased hard sphere fluids analysed with the Topological Cluster Classification.  We retrieve time series of $1000$ Monte-Carlo Sweeps (MCs) of the number of particles $N_i(t)$ or $n_i(t)=N_i(t)/N$ (the intensive concentration of structures of type $i$) for all the structures defined in the classification, an example of which is pictured in Fig.\ref{FigTimeSeries}. \ft{We note that, by definition, an individual particle may participate in more than a single motif. For example, it may be a constituent of two or more distinct pentagonal bypiramids. This is essential for the identification of larger structures (for instance, the 10B structure) and implies that the concentration $n_i(t)$ can in principle exceed unity for some motifs}. 

\begin{figure}
\begin{center}
\includegraphics[scale=1]{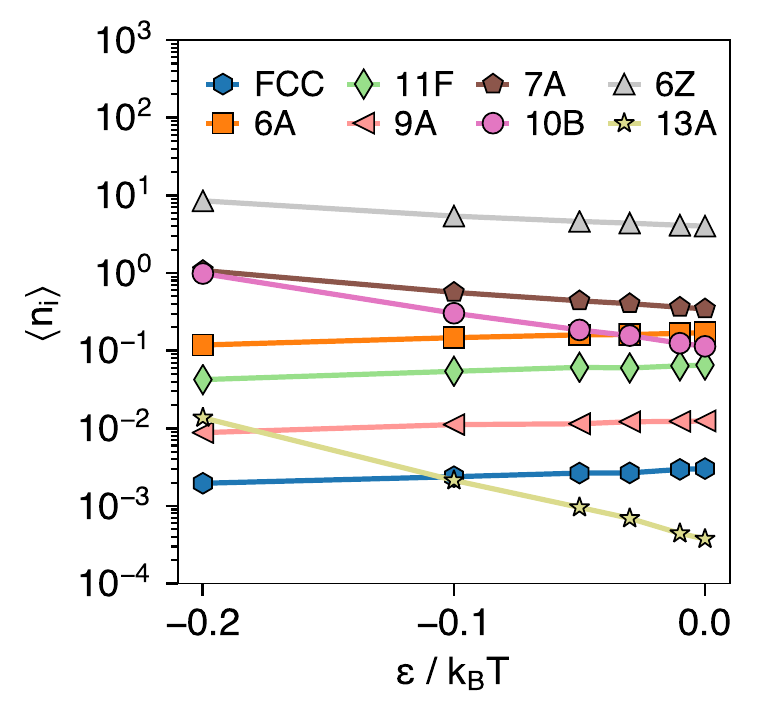}
\end{center}
\caption{\label{FigPopvsBiases}\ft{Average concentration} of detected structures, $\langle n_i\rangle$, in the system as a function of the bias $\varepsilon$ towards the pentagonal bipyramid ($7A$) structure. The packing fraction is constant at $\phi = 0.54$.}
\end{figure}

\begin{figure*}[t]
\centering
  \includegraphics[scale=1]{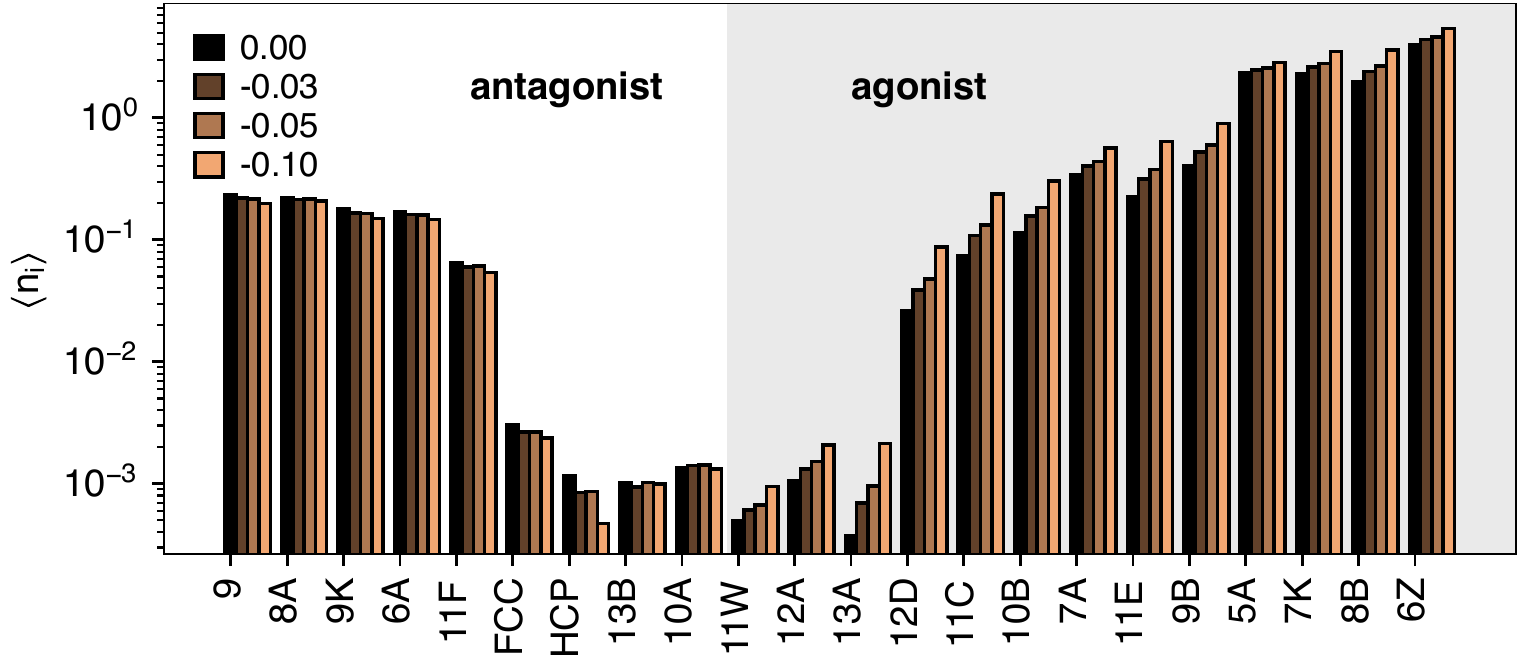}
  \caption{Effect of negative biases on the time-averaged concentrations $\langle n_i\rangle$ at packing fraction $\phi=0.54$ for structures in the Topological Cluster Classification with $\langle n_i \rangle > 10^{-4}$. Motifs that are agonist to $7A$ (shaded area) show an increase in concentration while the opposite occurs for the antagonist family of structures. }
  \label{FigAllNums}
\end{figure*}

Comparing the evolution of, for example, the 6Z and 6A structures with the 7A structure, we notice that while the former presents a very similar pattern to the pentagonal bipyramid ($n_{6Z}$ concentration increases as $n_{7A}$ increases), the other shows the opposite behaviour, suggesting that some structures are positively while others are negatively correlated to the five-fold symmetric structure. The time average $\langle n_i\rangle=\langle N_i\rangle/N$ for a selection of structures at packing fraction $\phi=0.54$ is plotted in Fig.~\ref{FigPopvsBiases} and shows that the concentrations of different structures differ of several orders of magnitude and have very different responses according to the change in the bias $\varepsilon$. A more complete picture for all the motifs with a significative average concentration $\langle n_i\rangle>10^{-4}$ is presented in Fig.~\ref{FigAllNums}. Unsurprisingly, small structures typically correspond to large concentrations while the opposite is true, in general, for structures composed of many particles. The largest structures such as the FCC, the HCP or 13A (i.e. icosahedral) motifs in the TCC comprise 13 particles and all have relatively small concentrations $n_i\sim 10^{-3}$. For very negative values of the bias $\varepsilon$,  7A structures are strongly favoured. This is clearly accompanied by the increase in the number of structures composed of 7A motifs such as 10B (termed \textit{defective icosahedron}) or 13A (the icosahedron) [see Fig.~\ref{FigFamilies} for three-dimensional rendering]. Correspondingly, the concentrations of structures related to four-fold symmetry, such as FCC or 11F, steadily drop at negative bias values.

To obtain the covariances we directly evaluate cross-correlations of the time-series at specific values of the packing fraction $\phi$ and bias $\varepsilon$. For any pair of structures $i,j$ in the classification, we define the matrix element
\begin{align}
	C_{i,j} (\phi, \varepsilon)& =N {\rm Cov} (n_i (\phi, \varepsilon), n_j(\phi, \varepsilon))\\
	& =\frac{N}{t_{\rm max}-1}\sum_{k=1}^{t_{\rm max}} (n_{i}(k)-\langle n_i\rangle) (n_{j}(k)-\langle n_j \rangle)
	\label{eq:covmat}
\end{align}
With such a definition, the covariance matrix is an intensive property
of the system. It should in principle depend on the packing fraction
$\phi$ and the bias $\varepsilon$. However, as we shall see in Section
\ref{sec:linear}, the knowledge of the covariance matrix in unbiased
conditions $C^0(\phi)=C(\phi, \varepsilon=0)$ is sufficient to
quantitatively predict changes in the structural properties of the
liquid.

\subsection{Structure of the covariance matrices}

\begin{figure*}[t]
\centering
\includegraphics[width=\textwidth]{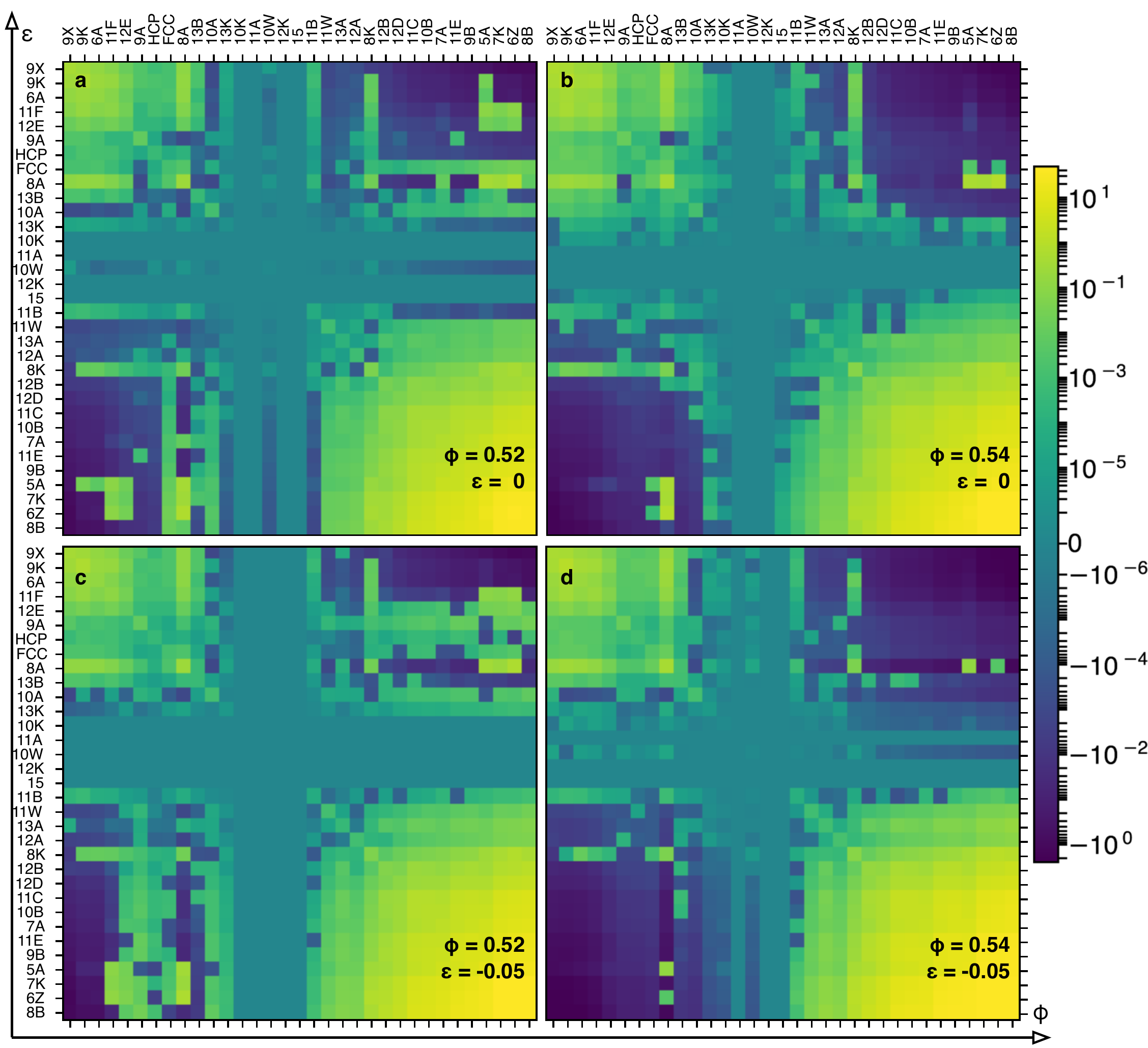}
\caption{\label{FigCovMats} Four examples of covariance matrices for different values of the bias $\varepsilon$ and packing fraction $\phi$: (a) $\varepsilon = 0$, $\phi=0.52$, (b) $\varepsilon = 0$, $\phi=0.54$; (c) $\varepsilon = -0.05$, $\phi=0.52$; (d) $\varepsilon = -0.05$, $\phi=0.54$. Very negative matrix elements are in blue while very positive matrix elements are in yellow. Structures are sorted according to the ascending order of their respective covariance with the pentagonal bipyramid $7A$ at a unbiased fixed state point $\phi=0.54, \varepsilon=0.0$. Notice the logarithmic color scale.}
\end{figure*}

We now discuss the properties of the covariance matrix $C(\phi, \varepsilon)$, obtained using Eq.~\ref{eq:covmat} over the set of $K=33$ structures defined in the Topological Cluster Classification that are composed of at least 5 particles. These structures include, for instance, the bi-tetrahedron (5A), the octahedron (6A), the 6-particle free energy minimum for six colloids with depleted mediated attractions (6Z), the pentagonal bipyramid (7A) as well as much larger structural motifs such as the defective icosahedron (10B), the icosahedron (13A) and crystalline motifs related to FCC (the 13-particle FCC motif) or HCP order (the 13-particle HCP cluster or the 11-particle 11F cluster). We show in Fig.~\ref{FigCovMats} four instances of the covariance matrix for different values of the packing fraction $\phi$ and the bias $\varepsilon$. These structures are sorted according to increasing covariance with the $7A$ structure for a reference case ($\phi=0.54$, $\varepsilon = 0$).

While there are some slight variations in the value of covariances, the overall structure of these matrices is essentially independent of the values of $\epsilon$ and $\phi$. A block structure with three groups of structures emerge: the left and right parts sets of structures exhibit strong positive correlation within each group and negative correlation to the opposite group, while the central part is a ``no-man's land'' with essentially zero covariances to all structures, including themselves. The rightmost group of structures contains 7A and all structures that correlate positively to it. Employing the language of Ronceray and Harrowell \cite{ronceray2017}, we term the structures $j$ with $C_{7A,j}>0$ \textit{agonist} to the pentagonal bipyramid $7A$ while those in the leftmost group, with $C_{7A,j}<0$, are \textit{antagonist} to $7A$. 

Going into further details, we observe that the largest covariances with 7A are $C_{7A, 6Z}$ and $C_{7A, 8B}$. The 8B structure is directly derived from the 7A bipyramid and has larger concentrations for combinatorial reasons (it corresponds to a 7A motif with an additional particle neighboring one of the two spindle particles, see Fig.\ref{FigFamilies}). The fact that the tetrahedral structure 6Z is a strong agonist is more surprising, as it does not contain any fivefold motif; we can rationalize its large magnitude by observing that it has an entropic advantage compared to, for example, the octahedron (6A) \cite{malins2009,meng2010}. The positive correlations revealed by the covariance analysis indicates that this structure overlaps well with 7A. These two examples illustrate a feature of the \textit{agonist} ($C_{7A,j}>0$) family: its members are either small structures with elementary tetrahedral order ($5A$, $6Z$, $7K$) or larger structures containing pentagonal rings ($10B$, $11C$,$11E$, $12B$, $12D$ and obviously $7A$ itself). This fact demonstrates that the covariance formalism is capable of detecting structural relationships between arbitrary motifs. 

Interestingly, the family of \textit{antagonist} structures ($C_{7A,j}<0$) displays positive mutual covariances $C_{ij}>0 : i,j \in \{\rm antagonist\}$, so that the top-left corner of the covariance matrix contains positive entries. Again, we can identify in the TCC definitions the geometric origin of these positive cross correlations: antagonist structures include the octahedron ($6A$), combinations of $6A$ such as $9K$, structures with pairs of square rings such as $9X$ and $9A$, or directly sections of crystalline cells such as the $11E$, $11F$ and $12E$ motifs and finally the HCP and FCC structures. This indicates that, within the Topological Cluster Classification, most of the antagonists to fivefold symmetry are of crystalline nature. The notable exception is provided by the $8A$ cluster (composed of very distorted pentagonal rings, strongly correlated with the $6Z$ tetrahedra and the $6A$ octahedron), and the $13B$ cluster (composed of two well aligned $7A$ clusters and hence mismatching both crystalline and icosahedral order).

We note that the while both the triangular bipyramid $5A$ and the octahedron $6A$ are originally both in the minimal energy structures of the HCP crystal in the case of other simple liquids such as the Lennard-Jones model \cite{malins2013tcc}, here they appear to play two different roles, the former correlating well with the emergence of pentagonal rings while the latter anticorrelates with it, promoting crystalline order instead. 

Finally, a \textit{no-man's land} of structures of effectively zero covariance separates the two families of agonist and antagonist structures. It includes structures such as $10W$ or $12K$ which have been defined in the TCC from minimum energy clusters of Lennard-Jones binary mixtures popular in the literature of the glass transition (the Wahnstr\"om \cite{wahnstrom1991} and the Kob-Andersen \cite{kob1994} respectively). The covariances for such clusters are null simply because the concentrations $n_{10W}$ and $n_{12K}$ are close to zero in the hard sphere liquid.

\subsection{Dependence on packing fraction and bias}

As the packing fraction or the bias vary, we move into different regions of the phase diagram in Fig.~\ref{FigPhaseDiagramSchematic}. Taking the high packing fraction unbiased point $\phi=0.54, \varepsilon=0$ (a metastable overcompressed liquid before nucleation occurs), we show in Fig.~\ref{FigCovMats} that the overall structure of the covariance matrix is broadly unchanged as we either reduce the packing fraction or bias the system to more negative values of $\varepsilon$, suppressing crystallization. 
We observe that, at the lower packing fraction, the antagonist family is restricted to a smaller number of structures, as large crystalline clusters such as $11F$, FCC or $12E$ present small covariances, due to the smaller concentrations of $n_{11F},n_{FCC}$ and $n_{12E}$ respectively. 
\begin{figure}[t]
\centering
  \includegraphics[scale=1]{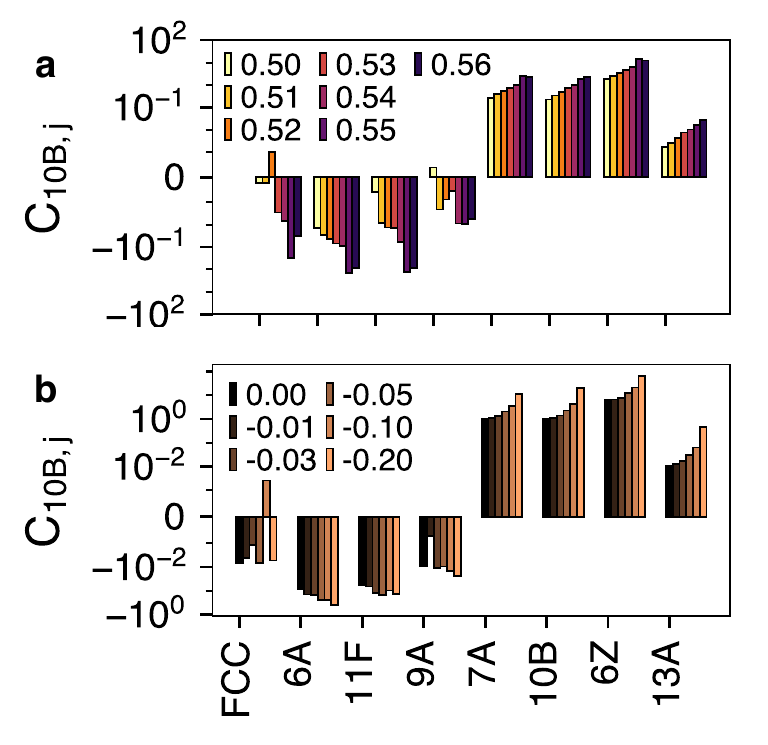}
  \caption{Example of the (a) packing fraction and (b) bias dependence of the covariance values between the agonist structure $10B$ and a selection of agonist and antagonist structures. In (a) the bias is $\varepsilon=0$ and in (b) the packing fraction is $\phi = 0.54$. }
  \label{FigRowCombined}
\end{figure}

In Fig.~\ref{FigRowCombined} we study the instructive case of the defective icosahedron (10B) structure and its covariances with notable members of the agonist and antagonist families. This is an \textit{agonist} structure to fivefold symmetry, as it is composed of three overlapping $7A$ motifs. The average concentration of this motif increases both as the packing fraction is increased and as the bias is more negative (see, for instance, Fig.\ref{FigPopvsBiases}). It is an important structure in hard sphere glasses as it dominates the free energy landscape in the metastable liquid branch at high densities \cite{royall2015,pinchaipat2017}. In Fig.~\ref{FigRowCombined}(a) we observe that increasing the packing fraction at zero bias leads to an increase in the magnitude of the covariance coefficients, which become more negative with the antagonist structures FCC, 6A, 11F and 9A and more positive with other agonist structures such as 7A, 6Z and the icosahedron 13A. This is an immediate consequence of the increase in the concentration of 10B at higher volume fractions compared to other structures, see Fig.~\ref{FigAllNums}. 

If we consider the dependence on the bias, Fig.~\ref{FigRowCombined}(b), we observe an analogous behaviour at constant packing fraction $\phi=0.54$. 
We also note that covariances with rare structures, such as the FCC crystalline motif, are very small and may flip sign with varying packing fraction/bias. This is the indication that more statistics (i.e. longer time series) are needed to more accurately estimate these covariances.

\section{Linear-response predictions}
\label{sec:linear}

The knowledge of the covariance matrix does not only provide insight
on the geometrical relationship between structures; it also allows us
to make quantitative predictions on the parameter dependence of the
liquid structure. Indeed, we can apply to our system the
fluctuation-response relation proposed by Ronceray and Harrowell in
\cite{ronceray_liquid_2016,ronceray2017} for on-lattice models, which
reads
\begin{equation}
  \langle n_i(\varepsilon) \rangle = \langle n_{i}^0 \rangle -\sum_{\textrm {structures }j }C_{i,j}\varepsilon_j, + O(\varepsilon^2)
  \label{eq:matrix_linresp}
\end{equation}
where $\varepsilon_j$ is the vector of energy biases associated to
each structure, $i$, such that the Hamiltonian is
$H = N \sum_i n_i \varepsilon_i$. The derivation remains correct in
our case, where the only nonzero bias is for the pentagonal bipyramid
$i=7A$. This results in a simple expression,
\begin{equation}
  \langle n_{i}( \phi, \varepsilon) \rangle = \langle n_{i}^0(\phi) \rangle - \varepsilon C_{i,7A}^0 +O(\varepsilon^2)
  \label{eq:linresp}
\end{equation}
where $n_{i}^0(\phi)$ is concentration of structure $i$ for the
unbiased system at packing fraction $\phi$, and $C_{7A,i}^0 $ is the
covariance matrix element between $i$ and $7A$ at packing fraction
$\phi$.

\begin{figure}[b]
  \centering
  \includegraphics[scale=1]{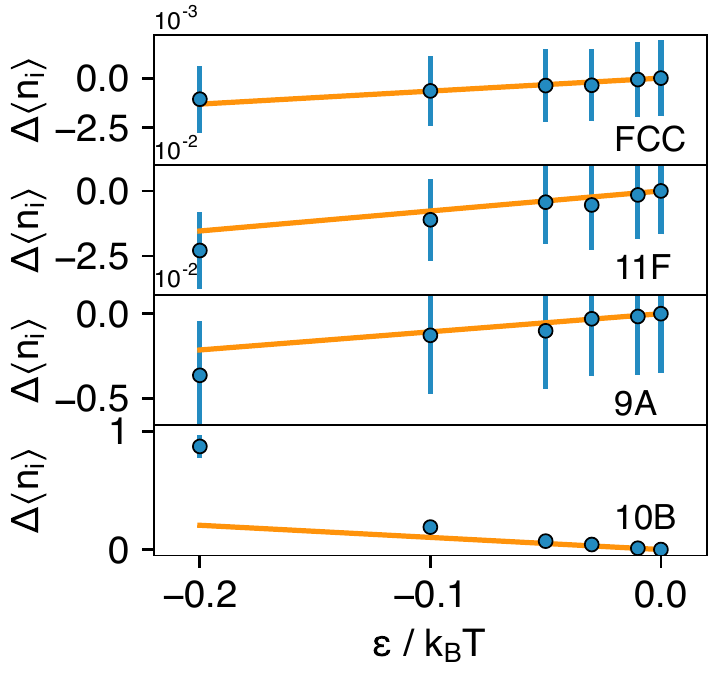}
  \caption{Tests of the linear response regime: the symbols represent the variations in concentration $\Delta n_i= n_i(\varepsilon) -n_i^0$ with vertical bars corresponding to one single standard deviation as computed from the Monte-Carlo trajectory. The straight orange lines are the predictions of Eq.~\ref{eq:linresp}, with covariances evaluated at $\varepsilon=0$. For all the plots, the packing fraction is $\phi=0.54$.}
\label{FigLinearResponse}
\end{figure}

Equation~\ref{eq:linresp} provides an exact prediction for the
first-order dependence of the structural composition of the liquid on
the applied structural bias. We demonstrate its validity in
Fig.~\ref{FigLinearResponse}, where we compare this linear-response
approximation and the measured change in concentrations
$\Delta n_i= \langle n_i(\varepsilon)\rangle -\langle n_i^0\rangle$
for four representative structures at fixed packing fraction
$\phi=0.54$: the $9A$, FCC and $11F$ (antagonist family), and $10B$
(agonist family). The linear prediction quantitatively captures the
bias dependency of the considered antagonist structures. For the
agonist structure $10B$, we observe higher-order deviations for large
biases $\varepsilon\leq-0.10$, with an accelerated accumulation of
these structures that is not captured by our linear theory. Note that
a similar trend is observed for agonist structures in lattice
models\cite{ronceray2017}.

Importantly, these results demonstrate that the accurate knowledge of
the covariance coefficient at a zero bias is sufficient to infer with
quantitative accuracy the structural changes in the system for biases
as large as $\varepsilon\approx\pm 0.1$.  This is not specific to
structure 7A. In principle, we could consider biasing the system
towards any single structure, or any weighted combination of
structures as in Equation~\ref{eq:matrix_linresp}: our approach
encompasses complex liquids described by an arbitrary set of biases
$\epsilon_i$, providing a predictive tool to quantitatively assess the
structure of any liquid at reasonably low value of the biases, or
equivalently at sufficiently high temperature. Beyond the linear
response regime, these results become quantitatively inaccurate, but
retain a qualitative pertinence: for instance, crystallization will
become essentially impossible if the concentrations of all four-fold
crystalline structures become too low.

\section{Conclusions}
\label{sec:conclusions}

Through the analysis of structural covariances in the biased hard sphere fluid we have shown that it is possible to understand how fivefold local order affects other competing motifs, such as those with four-fold symmetry which are related to crystalline order. We have discussed how covariances allow us to identify structural relationships between different motifs and we have illustrated how this applies to the particular case of the Topological Cluster Classification. Structural covariance reveals the existence of two main families of structures in the classification, pertaining to fivefold symmetric and crystal-like structures respectively. An interesting line of research would be to extend the approach to other classifications (such as the Voronoi indexing) and to compare different classification strategies according to the metric provided by the covariances.

In our study of the hard-sphere fluid we have found that the covariance approach is predictive in a wide range of bias values, estimating correctly, in the linear-response regime, structural changes for any of the structures classified in the Topological Cluster Classification. 

Our work demonstrates \ft{how an analysis based on structural covariances can be employed to investigate off-lattice models, providing a first proof of principle in the case of hard spheres. Other aspects of structural correlations in the fluid phase will deserve further study and comparison with the original on-lattice results. For example, in Ref.~\cite{ronceray2017} it has been shown that the so-called \textit{crystal affinity} $Q_{X}:= \partial n_X / \partial (1/T)$ can be expressed as $Q_X=-\sum_{j}C_{X,j}\varepsilon_{X,j}$ derived from the covariance coefficients between the crystalline motif $X$ and the remaining motifs. Remarkably, in Ref.~\cite{ronceray2017} the affinity $Q$ displays a characteristic anti-correlation with the crystallization times for the on-lattice systems. Understanding how this relation holds in the case of off-lattice models and how it depends on the specific identification of crystalline motifs (e.g. FCC, 11F or others such as bond order parameters\cite{steinhardt1983,lechner2009}) according to different structural descriptors will be the subject of further work.}



More generally, alternative routes to the calculation of the covariance matrix may provide efficient methods to estimate structural changes for a given set of structures: nonequilibrium protocols (such as shearing) are a potential avenue to measure structural couplings and covariances quickly and at a lower computational cost than biased Monte-Carlo. \ft{On the experimental side, since the knowledge of the local motifs is key to our approach, colloidal experiments (where the individual particle coordinates can be resolved) are most suitable for a test in the laboratory of the predictive power of the structural covariance analysis. However, since the covariances are computed between concentrations of different structures, spatial resolution is only necessary to identify chosen motifs. This means that as long as we are able to estimate local concentrations of particular motifs and preserve sample to sample variations, it is possible to compute covariances between distinct motifs even without the precise knowledge of all of the atomic positions. Advanced scattering techniques on molecular liquids (such as angstrom-beam electron diffraction \cite{dicicco2003,hirata2013}) may provide the route to measure such concentration and compute covariances between different sub-sampled regions of a dense, or supercooled, liquid.  }


\section*{Acknowledgements}
The authors are grateful to Jade Taffs for providing simulation data
and Joshua Robinson for his advice. FT, BMGDC and CPR thank the European
Research Council (ERC Consolidator Grant NANOPRS, project number
617266) for financial support. PR thanks the Bettencourt-Schueller
Foundation for their support.

\bibliography{allDropBox,additions.bib}
\bibliographystyle{unsrt}

\end{document}